\documentclass[10pt,twocolumn,journal]{IEEEtran}
\usepackage{times}
\usepackage{amsmath,amsfonts,amssymb}
\usepackage{latexsym}
\usepackage[ansinew]{inputenc}
\usepackage[dvips ]{graphicx}
\usepackage{color}
\usepackage{booktabs}
\usepackage{setspace}
\usepackage{listings}

\title{Interference Suppression in Multiuser Systems Based on Bidirectional Algorithms \vspace{-0.15em}}

\author{Patrick Clarke and Rodrigo C. de Lamare \vspace{-0.15em}
\thanks{Part of this manuscript was presented at the International
Symposium on Wireless Communications Systems (ISWCS) in 2013. Dr. P.
Clarke was with the Department of Electronics, University of York,
York Y010 5DD, United Kingdom, Prof. R. C. de Lamare is with
CETUC/PUC-RIO, 22453-900, Rio de Janeiro, Brazil and with the
Department of Electronics, University of York, York Y010 5DD, United
Kingdom. E-mails: rcdl500@ohm.york.ac.uk}}
\begin{document}
\maketitle
\begin{abstract}
This paper presents adaptive bidirectional minimum mean-square error
parameter estimation algorithms for fast-fading channels. The time
correlation between successive channel gains is exploited to improve
the estimation and tracking capabilities of adaptive algorithms and
provide robustness against time-varying channels. Bidirectional
normalized least mean-square and conjugate gradient algorithms are
devised along with adaptive mixing parameters that adjust to the
time-varying channel correlation properties. An analysis of the
proposed algorithms is provided along with a discussion of their
performance advantages. Simulations for an application to
interference suppression in multiuser DS-CDMA systems show the
advantages of the proposed algorithms.
\end{abstract}

\section{Introduction}
\label{5_1_Introdution}

Low-complexity reception and interference suppression are essential in
multiuser mobile systems if battery power is to be conserved, data-rates
improved and quality of service enhanced. Conventional adaptive schemes fulfil
many of these requirements and have been a significant focus of the research
literature
\cite{Multiuser_detection_Verdu,MMSE_int_supp_Honig,cdma_coop_mud_Huang,jpais,tds,sm_ce,jdpa,armo,itic,cdma_parallel_sic_lamare,mbdf,smadap}.
However, in time-varying fading channels commonly associated with mobile
systems, these adaptive techniques encounter tracking and convergence problems.
Optimum closed-form solutions can address these problems but their
computational complexity is high and CSI is required. Low-complexity adaptive
channel estimation can provide CSI but in highly dynamic channels tracking
problems exist due to their finite adaptation rate \cite{adapt_filt_Haykin}. An
alternative statistical approach is to obtain the correlation structures
required for optimal minimum mean-square error (MMSE) or least-squares (LS)
filtering \cite{MMSE_MUD_cdma_sadler, MMSE_noncoherent_int_supp_schoder}.
Although this relieves the tracking demands placed on the filtering process, in
a Rayleigh fading channel, a zero correlator is the result due to the
expectation of a Rayleigh fading coefficient, and therefore the
cross-correlation vector, equating to zero i.e. $E\left[h_{1}[n]\right]=0$ and
$E\left[b_{1}^{\ast}[n]\mathbf{r}[n]\right]=0$. In slowly fading channels this
problem may be overcome by using a time averaged approach where the averaging
period is equal to or less than the coherence time of the channel. However, in
fast fading channels an averaging period equal to the coherence time of the
channel is insufficient to overcome the effects of additive noise and
characterize the multiuser interference (MUI) \cite{Multiuser_detection_Verdu}.

Furthermore, the use of optimized convergence parameters such as step sizes and
forgetting factors into conventional adaptive algorithms extend their fading
range and lead to improved convergence and tracking performance \cite{smadap,
Extended_RLS_Haykin, variable_ff_rls_Wang, Variable_Step_size_lms_Kwong,
tracking_improv_rls_Toplis, gradient_variable_ff_rls_Leung,
variable_step_size_nlms_AP_Sayed, Gradient_based_var_step_lms_chambers}.
However, the stability of adaptive step-sizes and forgetting factors can be a
concern unless they are constrained to lie within a predefined region
\cite{Varibale_ss_lms_gelfand}. Other alternative schemes include those based
on processing the received data in subblocks
\cite{subblock,partialfft,mfsic_li,sicdma} and subspace algorithms
\cite{Honig02,sun,jiolms,jiols,jiomimo,jidf,saabf,barc}.In addition, the
fundamental problem of obtaining the unfaded symbols whilst suppressing MUI
remains. Consequently, the application of such algorithms is restricted to low
and moderate fading rates. The limitations of conventional estimation
approaches led to the development of methods that attempt to track the faded
symbol, such as the channel-compensated MMSE solution
\cite{CDMA_Flat_Rayleigh_Honig, adapt_mud_fading_channels_Poor}. This removes
the burden of fading coefficient estimation from the receive filter. However, a
secondary process is required to perform explicit estimation of the fading
coefficients in order to perform symbol estimation \cite{Diff_MMSE_Madhow}.

Approaches that avoid tracking and estimation of the fading coefficients were
proposed in \cite{CDMA_Serverly_Time_Varying_Channel_Madhow, Diff_MMSE_Madhow,
dynamic_fading_honig}. Although a channel might be highly time variant, two
adjacent fading coefficient will be similar and have a significant level of
correlation as studied in \cite{CDMA_Serverly_Time_Varying_Channel_Madhow,
Diff_MMSE_Madhow, dynamic_fading_honig}. These properties can then be exploited
to obtain a sequence of faded symbols where the primary purpose of the filter
is to suppress multiuser interference and track the ratio between successive
fading coefficients; thus, not burdening it with estimation of the fading
coefficients themselves. However, this scheme has a number of limitations
stemming from the use of only one correlation time instant and a single class
of adaptive algorithms.

{  In this work, a bidirectional MMSE based interference suppression scheme for
highly dynamic fading channels is presented. The non-zero correlation between
multiple time instants is exploited to improve the robustness, tracking and
convergence performance of existing MMSE schemes. Unlike existing adaptive
solutions \cite{CDMA_Serverly_Time_Varying_Channel_Madhow, Diff_MMSE_Madhow,
dynamic_fading_honig,smadap}, which do not fully exploit the fading correlation
between multiple successive time instants, the proposed bidirectional approach
exploits the correlation and adaptively weighs the output of the receive filter
in order to optimize the estimation performance. Normalized least-mean square
(NLMS) and conjugate gradient (CG) type algorithms are presented that overcome
a number of problems associated with applying the recursive least-squares (RLS)
algorithm to bidirectional problems. Novel mixing strategies that weigh the
contribution of the considered time instants and improve the convergence and
steady-state performance, increasing the robustness against the channel
discontinuities, are also presented. An analysis of the proposed schemes is
developed and establishes the mechanisms and factors behind their behaviour and
expected performance. The proposed schemes are applied to conventional
multiuser DS-CDMA \cite{MMSE_int_supp_Honig} and cooperative DS-CDMA systems
\cite{cdma_coop_mud_Huang,jpais} to assess their MUI suppression and tracking
capabilities. The application of the proposed scheme and algorithms to
multiple-antenna and multicarrier systems is also possible. Simulations show
that the algorithms improve upon existing schemes with minimal increase in
complexity.}

{ The main contributions of this work can be summarized as:
\begin{itemize}
\item{Bidirectional MMSE based interference suppression
scheme for highly dynamic fading channels.}
\item{Bidirectional adaptive parameter estimation algorithms based on NLMS and CG
techniques.}
\item{An analysis of the convergence and the computational complexity of the
proposed algorithms.}
\item{A study of the proposed and existing algorithms in DS-CDMA and cooperative
DS-CDMA multiuser systems.}
\end{itemize}
}

This paper is organized as follows. Section \ref{models} briefly details the
signal models of a conventional DS-CDMA system and a cooperative DS-CDMA
system. Section \ref{5_2_Proposed_Scheme} presents the proposed scheme and its
corresponding optimization problems and the motivation behind their
development. Switching and mixing strategies that optimize performance are
proposed and assessed in Section \ref{5_3_Switching Strategies}, followed by
the derivation of the proposed algorithms in Section \ref{5_4_Adaptive
Algorithms}. An analysis of the proposed algorithms is given in Section
\ref{5_5_Analysis}, whereas performance evaluation results are presented in
Section \ref{5_6_Simulations}. Conclusions are drawn in Section
\ref{5_7_Conclusions}.

{
\section{Signal Models}
\label{models}

In this section, we describe the signal models of a DS-CDMA system operating in
the uplink and a cooperative DS-CDMA system in the uplink equipped with relays
and the amplify-and-forward (AF) cooperation protocol. These systems are
employed for testing the proposed algorithms even though that extensions to
multiple-antenna and multi-carrier can also be considered with appropriate
modifications of the algorithms.

\subsection{DS-CDMA Signal Model}

We consider the uplink of a synchronous DS-CDMA system with $K$ users, $N$
chips per symbol and $L_p$ $(L_p<N)$ propagation paths for each link. We assume
that the delay is a multiple of the chip rate, the channel is constant during
each symbol interval and the spreading codes are repeated from symbol to
symbol. The received signal after filtering by a chip-pulse matched filter and
sampled at chip rate yields the $M$-dimensional received vector given by
\begin{equation}
\mathbf{r}[i] = A_{1}b_{1}[i]\mathbf{H}_{1}[i]\mathbf{c}_{1}[i]
+\underbrace{\sum^{K}_{k=2}A_{k}b_{k}[i]\mathbf{H}_{k}[i]\mathbf{c}_{k}[i]}_{\rm{MUI}}
+\boldsymbol{\eta}[i] + \mathbf{n}[i],
\end{equation}
where $M=N+L-1$, and $\mathbf{c}_{k}[i]$ and $A_{k}$ are the spreading sequence
and signal amplitude of the $k^{\mathrm{th}}$ user, respectively. The $M \times
N$ channel matrix with $L$ paths is given by $\mathbf{H}_{k}[i]$ for the
$k^{\mathrm{th}}$ user, the $M \times 1$ vector $\boldsymbol{\eta}[i]$
corresponds to the intersymbol interference and $\mathbf{n}[i]$ is the noise
vector. Conventional schemes use BPSK modulation and the differential and
bidirectional schemes employ differential BPSK where the sequence of data
symbols to be transmitted by the $k^{\mathrm{th}}$ user are given by $b_{k}[i]
= a_{k}[i]b_{k}[i-1]$ where $a_{k}[i]$ is the unmodulated baseband data.
Assuming that linear receive processing is adopted, the output of the receive
filter is given by
\begin{equation}
x[i] = {\mathbf w}^{H}[i]{\boldsymbol r}[i], \label{receiver1}
\end{equation}
where ${\mathbf w}[i]$ is an $M$-dimensional vector that corresponds to the
receive filter.

\subsection{Cooperative DS-CDMA Signal Model}

\begin{figure}
 \centering
\includegraphics[width=0.9\columnwidth]{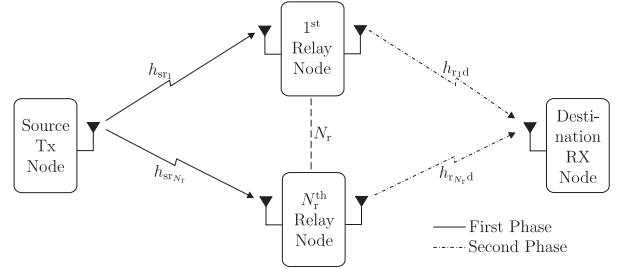}
\vspace{-0.5em}
  \caption{Cooperative DS-CDMA System Model}
  \label{fig:Cooperative_CDMA_System_Model}
\end{figure}

We also consider the uplink of a cooperative DS-CDMA system with $K$ users,
$N_r$ relays, $N$ chips per symbol and $L_p$ $(L_p<N)$ propagation paths for
each link. The system shown in \ref{fig:Cooperative_CDMA_System_Model} is
equipped with an AF protocol at each relay. The received signals at the
$n^{\mathrm{th}}$ relay and the destination nodes are filtered by a chip-pulse
matched filter, sampled at chip rate to obtain sufficient statistics and
organized into $M \times1$ vectors as described by
\begin{equation}
\mathbf{r}_{\mathrm{sr}_{n}}[i] = \sum^{K}_{k=1}
a_{\mathrm{s}_{k}}[i]b_{k}[i]h_{\mathrm{sr}_{n}}[i]\mathbf{c}_{k}[i] +
\mathbf{n}_{\mathrm{r}_{n}}[i],
\end{equation}
\begin{equation}
\mathbf{r}_{\mathrm{rd}}[i] =
\sum^{N_{\mathrm{r}}}_{n=1}a_{\mathrm{r}_{n}}[i]h_{\mathrm{r}_{n}\mathrm{d}}[i]
\mathbf{r}_{\mathrm{sr}_{n}}[i] + \mathbf{n}_{\mathrm{d}}[i]
\end{equation}
and
\begin{equation}
\mathbf{r}_{\mathrm{rd}}[i]
=\sum^{N_{\mathrm{r}}}_{n=1}\sum^{K}_{k=1}a_{\mathrm{s}_{k}}[i]
a_{\mathrm{r}_{n}}[i] h_{\mathrm{sr}_{n}}[i] h_{\mathrm{r}_{n}\mathrm{d}}[i]
\mathbf{c}_{k}[i] b_{k}[i] + \sum^{N_{\mathrm{r}}}_{n=1} a_{\mathrm{r}_{n}}[i]
h_{\mathrm{r}_{n}\mathrm{d}}[i]\mathbf{n}_{\mathrm{r}_{n}}[i] +
\mathbf{n}_{\mathrm{d}}[i].
\end{equation}
where $h_{\mathrm{sr}_{n}}[i]$ and $h_{\mathrm{r}_{n}\mathrm{d}}[i]$ are the
channel fading channel coefficients between the source and the
$n^{\mathrm{th}}$ relay, and the $n^{\mathrm{th}}$ relay and the destination,
respectively, and $\mathbf{n}_{\mathrm{r}_{n}}[i]$ and
$\mathbf{n}_{\mathrm{d}}[i]$ are additive white Gaussian noise vectors at the
relays and the destination, respectively.

The received data is processed by a linear receive filter, which produces the
output given by
\begin{equation}
x[i] = {\mathbf w}^{H}[i]{\boldsymbol r}_{\rm rd}[i], \label{receiver2}
\end{equation}
where ${\mathbf w}[i]$ is an $M$-dimensional vector that corresponds to the
receive filter for the cooperative system.}

\section{Proposed Bidirectional Scheme}
\label{5_2_Proposed_Scheme}

Adaptive parameter estimation has two primary objectives: estimation and
tracking of the desired parameters. When applied to multiuser wireless systems,
these translate into recovery of the desired symbol, tracking of channel
variations and suppression of MUI. However, in fast fading channels these
objectives place unrealistic demands on conventional filtering and estimation
schemes. Differential techniques reduce these demands by relieving adaptive
receivers from the task of tracking fading coefficients
\cite{Diff_MMSE_Madhow}. This is achieved by posing an optimization problem
where the ratio between two successive received samples is the quantity to be
tracked. Such an approach is enabled by the presumption that, although the
fading is fast, there is correlation between the adjacent channel samples as
described by
\begin{equation}
f_{1}[i]=E\left[h_{1}[i]h_{1}^{\ast}[i+1]\right]\geq 0, \\
\end{equation}
where $h_{1}[i]$ is the channel coefficient of the desired user. The
interference suppression of the resulting receive filter is improved in fast
fading environments compared to conventional adaptive receivers but only the
ratio of adjacent fading samples is obtained. Consequently, differential MMSE
schemes are suited to differential modulation where the ratio between adjacent
symbols is the data carrying mechanism.

\begin{figure}
\centering
  {\includegraphics[width=0.7\columnwidth]{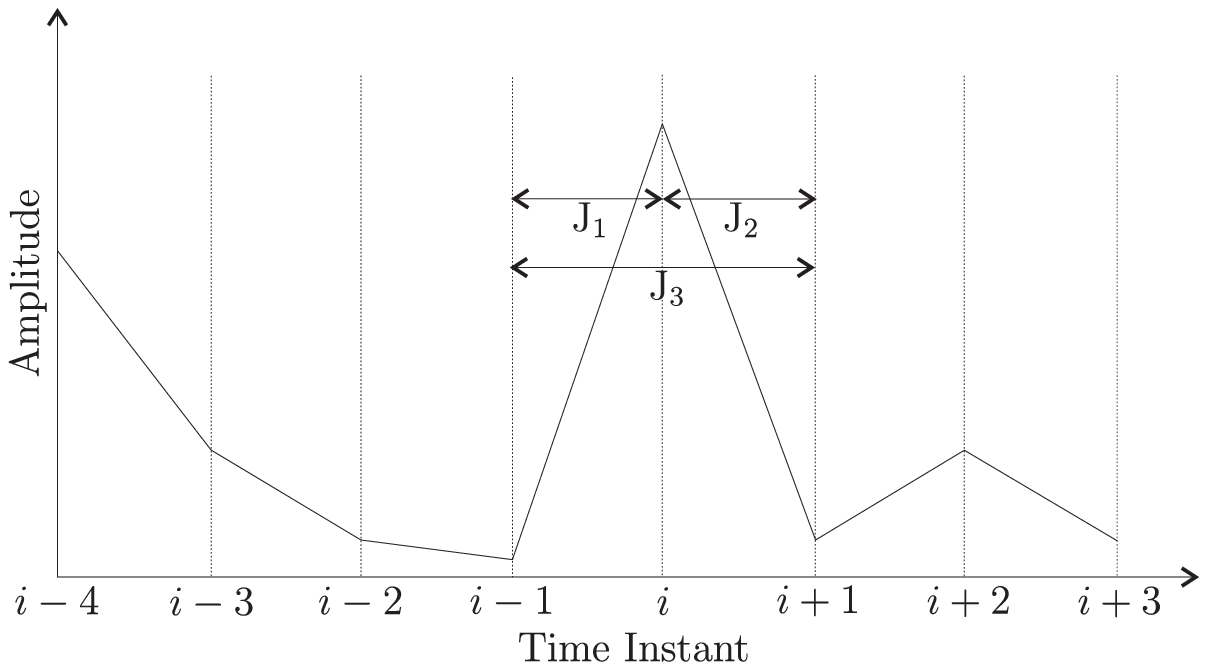}
  \label{Fig:Uncorrelated_fading_channel}}\\
  {\includegraphics[width=0.7\columnwidth]{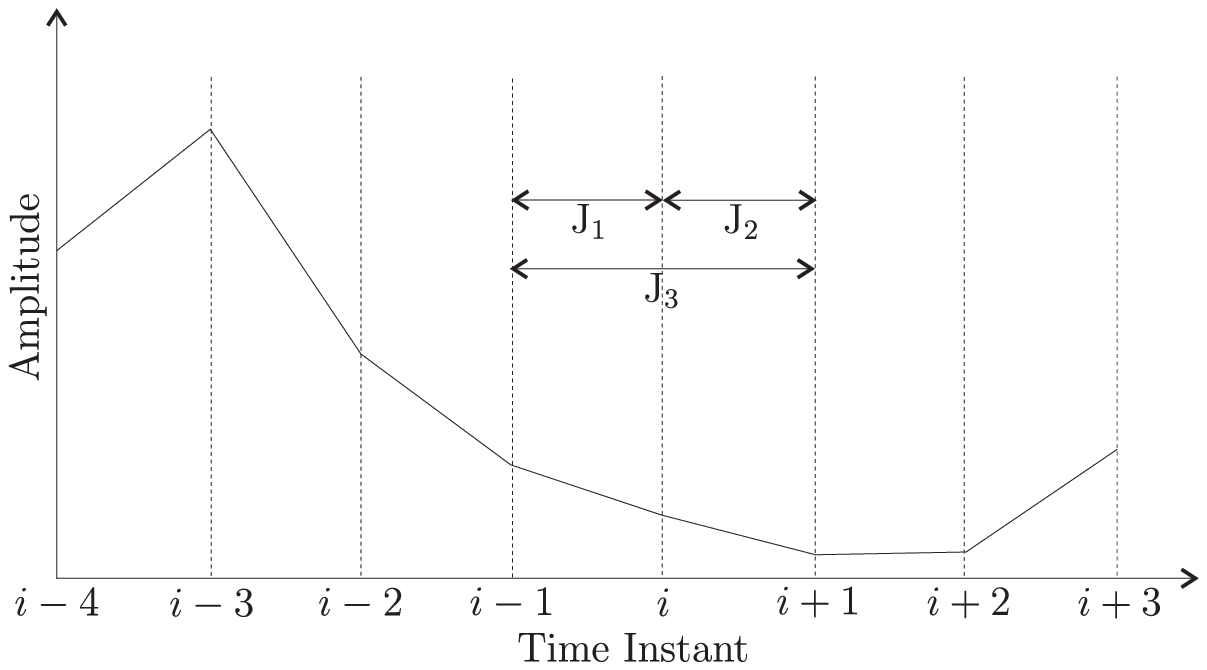}
  \label{Fig:Correlated_fading_channel}}\\
\caption{Fading channels: an uncorrelated fading channel on the top
figure and a correlated fading channel on the bottom figure.}
\label{Fig:Fading_channels}
\end{figure}

However, limiting the optimization to two adjacent samples exposes these
processes to the negative effects of uncorrelated samples
\begin{equation}
E\left[h_{1}[i]h_{1}^{\ast}[i+1]\right]\approx 0,
\end{equation}
but also does not exploit the correlation that may be present
between two or more adjacent samples, i.e.,
\begin{equation}
\begin{array}{l}
f_{2}[i]=E\left[h_{1}[i]h_{1}^{\ast}[i-1]\right]>0\\
f_{3}[i]=E\left[h_{1}[i+1]h_{1}^{\ast}[i-1]\right]>0.\\
\end{array}
\end{equation}
In order to address these weaknesses, we propose a bidirectional MSE
cost function based on multiple adjacent samples so that the number
of channel scenarios under which the differential MMSE performs
beneficial adaptation is substantially increased. Termed the
bidirectional MMSE, due to the use of multiple time instants, the
motivation behind this proposition is illustrated by the plots of
fading/channel coefficients in Fig. \ref{Fig:Fading_channels}, where
$\mathrm{J}_{1}$ represents the $2$ sample differential MMSE. There
is a low level of correlation present between samples $i$ and $i-1$,
thus any adaptation of the receive filter will bring little benefit.
However, the proposed scheme for $3$ time instants operates over
$\mathrm{J}_{1}$,$\mathrm{J}_{2}$ and $\mathrm{J}_{3}$; therefore,
it can exploit the correlation between $i+1$ and $i-1$ and past
data. Figure \ref{Fig:Fading_channels} gives an example of a channel
where there is a significant level of correlation between samples.

The proposed bidirectional scheme can be expressed in the form of an
an optimization problem as described by
\begin{equation}
\begin{array}{lll}
\mathbf{w}_o[i] = \underset{\mathbf{w}[i]}{\mathrm{arg \; min}}  &
E\left[\left|b[i]\mathbf{w}^{H}[i]
\mathbf{r}[i-1]-b[i-1]\mathbf{w}^{H}[i]\mathbf{r}[i]\right|^{2}\right.\\
&\hspace{3.2cm}\vdots\\
&+\left|b[i]\mathbf{w}^{H}[i]\mathbf{r}[i-(D-1)]-b[i-(D-1)]\mathbf{w}^{H}[i]\mathbf{r}[i]\right|^{2}\\
&\\
&+\left|b[i-1]\mathbf{w}^{H}[i]\mathbf{r}[i-2]-b[i-2]\mathbf{w}^{H}[i]\mathbf{r}[i-1]\right|^{2}\\
&\hspace{3.2cm}\vdots \\
&+\left|b[i-1]\mathbf{w}^{H}[i]\mathbf{r}[i-(D-1)]-b[i-(D-1)]\mathbf{w}^{H}[i]\mathbf{r}[i-1]\right|^{2}\\
&\\
&+\left.\left|b[i-(D-2)]\mathbf{w}^{H}[i]\mathbf{r}[i-(D-1)]-b[i-(D-1)]\mathbf{w}^{H}[i]
\mathbf{r}[i-(d-2)]\right|^{2}\right]\\
\end{array},
\label{eq:Generalised_Bidirectional_Cost_Function}
\end{equation}
where $D$ denotes the number of considered time instants.
Introducing summations into
(\ref{eq:Generalised_Bidirectional_Cost_Function}) yields a more
concise form
\begin{equation}
\mathbf{w}_o[i] = \underset{\mathbf{w}[i]}{\mathrm{arg \; min}}
E\left[\displaystyle\sum^{D-2}_{d=0}\sum^{D-1}_{l=d+1}
\left|b[i-d]\mathbf{w}^{H}[i]\mathbf{r}[i-l]-b[i-l]
\mathbf{w}^{H}[i]\mathbf{r}[i-d]\right|^{2}\right],
\label{eq:Generalised_Bidirectional_Cost_Function_Sum}
\end{equation}
where an output power constraint is required to avoid the trivial
all-zero receive filter solution
\begin{equation}
E\left[\left|\mathbf{w}^{H}[i]\mathbf{r}[i]\right|^{2}\right]=1.
\label{eq:Non_Zero_Constraint}
\end{equation}
{  Although the existing differential scheme operates over $2$ correlated
samples, the proposed scheme is able to exploit the additional correlation
present between multiple adjacent samples.  Moreover, it is also possible to
obtain further gain by weighting the correlation between multiple adjacent
samples. However, the benefit of using multiple time instant is dependent on
the fading rate of the channel and the related correlation of the channel
coefficients. We have investigated the use of multiple time instants and it
turns out that a scheme which exploits $3$ adjacent samples captures most of
the performance benefits. In particular, we have tested the proposed
bidirectional scheme and algorithms with various values of adjacent time
instants (between 4 and 8) and verified that exploiting extra time instants
above 3 does not yield significant gains. In fact, the number of time instants
is a parameter to be chosen by the designer.}

The optimization problem of the proposed scheme for $3$ time
instants is given by
\begin{equation}
\begin{array}{lllr}
\mathbf{w}_o[i] = \underset{\mathbf{w}[i]}{\mathrm{arg \; min}}  &
E\left[\left|b[i]\mathbf{w}^{H}[i]\mathbf{r}[i-1]-b[i-1]\mathbf{w}^{H}[i]\mathbf{r}[i]
\right|^{2}\right. &(\mathrm{J}_{1})\\
 &+\left|b[i]\mathbf{w}^{H}[i]\mathbf{r}[i-2]-b[i-2]\mathbf{w}^{H}[i]
\mathbf{r}[i] \right|^{2}&(\mathrm{J}_{2})\\
 &+\left.\left|b[i-1]\mathbf{w}^{H}[i]\mathbf{r}[i-2]-
b[i-2]\mathbf{w}^{H}\mathbf{r}[i-1] \right|^{2}\right]&(\mathrm{J}_{3})\\
\label{eq:Bidirectional_Cost_Function}
\end{array},
\end{equation}
where $\mathrm{J}_{1} - \mathrm{J}_{3}$ equate to those of Fig.
\ref{Fig:Fading_channels}, and the time instants of interest have been altered
to avoid the use of future samples. In addition to
(\ref{eq:Bidirectional_Cost_Function}), an output power constraint is again
required to avoid an all-zero trivial solution as given by
\begin{equation}
E\left[\left|\mathbf{w}^{H}[i]\mathbf{r}[i]\right|^{2}\right]=1.
\end{equation}
{  In what follows, we describe switching and weighting strategies to optimize
the proposed scheme and obtain further performance gain.}

\section{Switching and Weighting Strategies}
\label{5_3_Switching Strategies}

The advantages of a bidirectional scheme operating over 3 time or more time
instants have been verified in our studies. However, the performance of the
scheme may be degraded when received vectors based on uncorrelated fading
coefficients are employed in the update of the receive filter. This is
particulary evident from the example with an uncorrelated channel illustrated
in Fig. \ref{Fig:Fading_channels}, where the contribution to the cost function
represented by $\mathrm{J}_{3}$ is unlikely to aid the accurate adaptation of
$\mathbf{w}[i]$. To avoid this we introduce a set of switching or mixing
parameters that determine the weighting of the $D$ constituent elements of the
bidirectional cost function. The proposed generalized bidirectional cost
function with weighting factors is described by
\begin{equation}
\mathbf{w}_o[i] = \underset{\mathbf{w}[i]}{\mathrm{arg \; min}}
E\left[\displaystyle\sum^{D-2}_{d=0}\sum^{D-1}_{l=d+1}\rho_{n}[i]\left|b[i-d]\mathbf{w}^{H}[i]
\mathbf{r}[i-l]-b[i-l]\mathbf{w}^{H}[i]\mathbf{r}[i-d]\right|^{2}\right],
\label{eq:Generalised_Bidirectional_Cost_Function_Sum_Mixing}
\end{equation}
where $n=d(D-3)+l+1$. However, we again focus on the case with $D=3$
in the remainder of this work. With these modifications the proposed
bidirectional MSE cost function with $3$-time instand and weighting
factors is given by
\begin{equation}
\begin{array}{lllr}
\mathbf{w}_o[i] = \underset{\mathbf{w}[i]}{\mathrm{arg \; min}}
&E\left[\rho_{1}[i]\left|b[i]\mathbf{w}^{H}[i] \mathbf{r}[i-1]-
b[i-1]\mathbf{w}^{H}[i]\mathbf{r}[i] \right|^{2}\right.&(\mathrm{J}_{1})\\
&+\rho_{2}[i]\left|b[i]\mathbf{w}^{H}[i]\mathbf{r}[i-2]-b[i-2]
\mathbf{w}^{H}[i] \mathbf{r}[i] \right|^{2}&(\mathrm{J}_{2})\\
&+\left.\rho_{3}[i]\left|b[i-1]\mathbf{w}^{H}[i] \mathbf{r}[i-2]
-b[i-2]\mathbf{w}^{H}[i]\mathbf{r}[i-1] \right|^{2}\right]&(\mathrm{J}_{3})\\
\label{eq:Bidirectional_Cost_Function_Switching}
\end{array},
\end{equation}
where $0 \leq \rho_n \leq3$ for $n=1,2,3$ are the weighting factors.

The determination of the receive vector samples that correspond to the
scenarios depicted in Figure \ref{Fig:Fading_channels} is essential if correct
optimization of the $\rho$ is to be achieved. The use of CSI to achieve this
would be an effective but impractical solution due to the difficulty in
obtaining CSI; consequently, other methods must be sought. In this section, we
propose the use of two alternative metrics: the signal power differential after
interference suppression between the considered time instants, and the error
between the considered time instants.

Firstly, we consider a switching scheme where $\rho_{1-3} = [0,1]$
are determined at each time instant based on the following
post-filtering power differential metrics:
\begin{equation}
\begin{array}{l}
P_{1}[i]=\left\vert\mathbf{w}[i]^{H}\mathbf{r}[i]\right\vert^{2}-\left\vert\mathbf{w}[i]^{H}\mathbf{r}[i-1]\right\vert^{2}\\
P_{2}[i]=\left\vert\mathbf{w}[i]^{H}\mathbf{r}[i]\right\vert^{2}-\left\vert\mathbf{w}[i]^{H}\mathbf{r}[i-2]\right\vert^{2}\\
P_{3}[i]=\left\vert\mathbf{w}[i]^{H}\mathbf{r}[i-1]\right\vert^{2}-\left\vert\mathbf{w}[i]^{H}\mathbf{r}[i-2]\right\vert^{2}.\\
\end{array}
\end{equation}
If the power difference for each of $\mathrm{J}_{1-3}$ exceeds a
predefined threshold, the corresponding $\rho$ is set to zero;
therefore, removing the corresponding element of the cost function
from the adaptation process at that time instant. For highly dynamic
channels one requires an adaptive threshold which is able to track
the changes in the system and determine appropriate time instants
based on successive samples. Consequently, for each $\rho_{n}$ a
threshold, $T_{n}[i]$, related to a time-averaged, windowed,
root-mean-square of the relevant differential power is used. The
value of $\rho_{n}$ in then determined in the following manner:
\begin{equation}
\rho_{n}[i] = \left\{
\begin{array}{ll}
0& \mathrm{if}~ P_{n}[i] \geq T_{n}[i]\\
1& \mathrm{otherwise}\\
\end{array},
\right.
\label{eq:swithcing_factors}
\end{equation}
where
\begin{equation}
T_{n}[i]= \nu\left[\lambda_{P}P_{n_{RMS}}[i] + (1-\lambda_{P})P_{n_{RMS}}[i]\right],
\end{equation}
\begin{equation}
P_{n_{RMS}}[i]=\sqrt{\frac{1}{m-1}\sum^{i}_{l=i-m}P_{n}[l]^{2}},
\end{equation}
{  and $\nu$ is a positive user defined constant greater than unity that scales
the threshold. The threshold $\nu$ is set with the help of computer experiments
in a similar way as the step size of the NLMS algorithm is tuned. The aim is to
scale the threshold such that it will be used to inform the algorithm about the
relevant differential power which should be used. }

Although the current sample corresponding to $\mathrm{J}_{n}$ may
bring little benefit in terms of adaptation, this does not indicate
that all previous cost function elements corresponding to
$\mathrm{J}_{n}$ should be discarded. An alternative approach is to
use a set of convex mixing parameters that are not restricted to 1
or 0. This allows each element of the cost function to be more
precisely weighted based on its previous and current values.
However, the setting of these mixing parameters is once again
problematic if they are fixed. Accordingly, an adaptive
implementation that can take account of the time-varying channels
and previous values which continue to have an impact on the
adaptation of the filter is sought. The errors extracted from the
cost function (\ref{eq:Bidirectional_Cost_Function_Switching}) are
chosen as the metric for developing algorithms. These provide an
input to the weighting factor calculation process that is directly
related to the cost function of
(\ref{eq:Bidirectional_Cost_Function_Switching}). The time-varying
mixing factors are given by
\begin{equation}
\rho_{n}[i] = \lambda_{e}\rho_{n}[i-1] + (1-\lambda_{e})\frac{e_{T}[i]-|e_{n}[i]|}{e_{T}[i]}
\label{eq:mixing_factors}
\end{equation}
where
\begin{equation}
e_{T}[i] = |e_{1}[i]|+|e_{2}[i]|+|e_{3}[i]|,
\end{equation}
and the individual error terms are calculated as
\begin{equation}
\begin{array}{l}
e_{1}[i] = b[i]\mathbf{w}^{H}[i-1]\mathbf{r}[i-1]-b[i-1]\mathbf{w}^{H}[i-1]\mathbf{r}[i]\\
e_{2}[i] = b[i]\mathbf{w}^{H}[i-1]\mathbf{r}[i-2]-b[i-2]\mathbf{w}^{H}[i-1]\mathbf{r}[i]\\
e_{3}[i] = b[i-1]\mathbf{w}^{H}[i-1]\mathbf{r}[i-2]-b[i-2]\mathbf{w}^{H}[i-1]\mathbf{r}[i-1].\\
\end{array}
\end{equation}
The forgetting factor, $0\leq \lambda_{e}\leq1$, is user defined and, along
with normalization by the total error, $e_{T}[i]$, and
$\displaystyle{\sum^{3}_{n=1}}\rho_{n}[0]=1$,  ensures
$\displaystyle{\sum^{3}_{n=1}}\rho_{n}[i]=1$ and a convex combination at each
time instant.

\section{Adaptive Algorithms}
\label{5_4_Adaptive Algorithms}

In order to devise low-complexity adaptive algorithms based on the
proposed bidirectional schemes, we consider the minimization of the
cost function given by
\begin{equation}
\begin{array}{ll}
C(\mathbf{w}[i]) =  &E\left[\left|b[i]\mathbf{w}^{H}[i-1]\mathbf{r}[i-1]-b[i-1]\mathbf{w}^{H}[i]\mathbf{r}[i] \right|^{2}\right.\\
& +\left|b[i]\mathbf{w}^{H}[i-2]\mathbf{r}[i-2]-b[i-2]\mathbf{w}^{H}[i]\mathbf{r}[i] \right|^{2}\\
& +\left.\left|b[i-1]\mathbf{w}^{H}[i-2]\mathbf{r}[i-2]-b[i-2]\mathbf{w}^{H}[i]\mathbf{r}[i-1] \right|^{2}\right]\\
&\\
\mathrm{subject\,to}& E\left[\left|\mathbf{w}[i]^{H}\mathbf{r}[i]\right|^{2}\right]=1.
\end{array}
\label{eq:Bidirectional_Cost_Function_AA}
\end{equation}
This cost function then forms the basis of the adaptive algorithms derived in
this section. However, to reduce the complexity of the derivations, enforcement
of the non-zero constraint is not included and instead enforced in a stochastic
manner at each time instant after the adaptation step is complete
\cite{Diff_MMSE_Madhow}.


\subsection{Normalized Least-Mean Square Algorithm}
\label{5_4_1_Stochastic Gradient}

{We begin with the low-complexity NLMS implementation that employs an
instantaneous gradient in a steepest descent framework.} Firstly, the
instantaneous gradient of (\ref{eq:Bidirectional_Cost_Function_AA}) is taken
with respect to $\mathbf{w}^{*}[i]$, yielding
\begin{equation}
\begin{array}{ll}
\nabla_{\mathbf{w}^{*}[i]}C(\mathbf{w}[i])=&-b[i-1]\mathbf{r}[i]
(b[i]\mathbf{w}^{H}[i-1]\mathbf{r}[i-1]-b[i-1]\mathbf{w}^{H}[i]\mathbf{r}[i])^{H}\\
&-b[i-2]\mathbf{r}[i](b[i]\mathbf{w}^{H}[i-2]\mathbf{r}[i-2]-b[i-2]\mathbf{w}^{H}[i]\mathbf{r}[i])^{H}\\
&-b[i-2]\mathbf{r}[i-1](b[i-1]\mathbf{w}^{H}[i-2]\mathbf{r}[i-2]-b[i-2]\mathbf{w}^{H}[i]\mathbf{r}[i-1] )^{H}
\end{array}
\label{eq:NMLS_Gradient}
\end{equation}
At this point, in order to improve the convergence performance of
the NLMS algorithm, the bracketed error terms of
(\ref{eq:NMLS_Gradient}) are modified by replacing the receive
filters with the most recently calculated one, $\mathbf{w}[i-1]$.
The resulting gradient expression is given by
\begin{equation}
\begin{array}{ll}
\nabla_{\mathbf{w}^{*}[i]}C(\mathbf{w}[i])=&-b[i-1]\mathbf{r}[i]
\underbrace{\left(b[i]\mathbf{w}^{H}[i-1]\mathbf{r}[i-1]-b[i-1]\mathbf{w}^{H}[i-1]\mathbf{r}[i]\right)^{H}}_{e_{1}[i]}\\
&-b[i-2]\mathbf{r}[i]\underbrace{\left(b[i]\mathbf{w}^{H}[i-1]
\mathbf{r}[i-2]-b[i-2]\mathbf{w}^{H}[i-1]\mathbf{r}[i]\right)^{H}}_{e_{2}[i]}\\
&-b[i-2]\mathbf{r}[i-1]\underbrace{\left(b[i-1]\mathbf{w}^{H}[i-1]\mathbf{r}[i-2]-
b[i-2]\mathbf{w}^{H}[i-1]\mathbf{r}[i-1]\right)^{H}}_{e_{3}[i]}
\end{array}.
\label{eq:NMLS_Gradient_Upadted_w}
\end{equation}
Placing the above gradient expression in the steepest descent update
recursion, we obtain
\begin{equation}
\mathbf{w}[i] = \mathbf{w}[i-1] +
\frac{\mu}{M[i]|\mathbf{w}^{H}[i-1]\mathbf{r}[i-1]|}\left[b[i-1]
\mathbf{r}[i]e_{1}^{\ast}[i] + b[i-2]\mathbf{r}[i]e_{2}^{\ast}[i] +
b[i-2]\mathbf{r}[i-1]e_{3}^{\ast}[i]\right], \label{eq:NLMS_Update}
\end{equation}
where $\mu$ is the step-size and the normalization factor, $M[i]$, is given by
\begin{equation}
M[i] = \lambda_{M}M[i-1] + (1-\lambda_{M})\mathbf{r}^{H}[i]\mathbf{r}[i],
\end{equation}
where $\lambda_{M}$ is an exponential forgetting factor
\cite{Diff_MMSE_Madhow}. The enforcement of the constraint is performed by the
denominator of (\ref{eq:NLMS_Update}) which ensures that the receive filter
$\mathbf{w}[i]$ does not tend towards a zero correlator as the adaptation
progresses.

The incorporation of the variable switching and mixing factors of Section
\ref{5_3_Switching Strategies} has the potential to improve the performance of
the above algorithm by {  optimizing} the weighting of the error terms of
(\ref{eq:NMLS_Gradient_Upadted_w}). Integration of the factors given by
(\ref{eq:swithcing_factors}) and (\ref{eq:mixing_factors}) yields
\begin{equation}
\mathbf{w}[i] = \mathbf{w}[i-1] +
\frac{\mu}{M[i]|\mathbf{w}^{H}[i-1]\mathbf{r}[i-1]|}
\left[\rho_{1}[i]b[i-1]\mathbf{r}[i]e_{1}[i] +
\rho_{2}[i]b[i-2]\mathbf{r}[i]e_{2}[i] +
\rho_{3}[i]b[i-2]\mathbf{r}[i-1]e_{3}[i]\right] \label{eq:NLMS_Update_mixing}
\end{equation}
as the receive filter update equation.

\subsection{Least Squares Algorithm}
\label{5_4_2_Least Squares}

To achieve faster convergence and increased robustness to fading we
now pursue a LS based solution. Firstly, the bidirectional cost
function of (\ref{eq:Bidirectional_Cost_Function_AA}) is cast as an
LS problem by replacing the expected value with a weighted
summation, as described by
\begin{equation}
\begin{array}{llll}
C_{\rm LS}(\mathbf{w}[i]) & = \displaystyle{\sum^{i}_{l=1}}&\lambda^{i-l}&\left[\left|b[i]\mathbf{w}^{H}[i-1]\mathbf{r}[i-1]-b[i-1]\mathbf{w}^{H}[i]\mathbf{r}[i] \right|^{2}+\right.\\
& & & \left|b[i]\mathbf{w}^{H}[i-2]\mathbf{r}[i-2]-b[i-2]\mathbf{w}^{H}[i]\mathbf{r}[i] \right|^{2}+\\
& & & \left.\left|b[i-1]\mathbf{w}^{H}[i-2]\mathbf{r}[i-2]-b[i-2]\mathbf{w}^{H}[i-1]\mathbf{r}[i-1] \right|^{2}\right]\\
\end{array},
\label{eq:LS_Bidirect_Cost_Function}
\end{equation}
where $\lambda$ is an exponential forgetting factor. Proceeding as with the
conventional LS derivation, and modifying the equivalent error terms in a
similar manner to as in (\ref{eq:NMLS_Gradient_Upadted_w}), we arrive at the
following expressions for the component autocorrelation matrices:
\begin{equation}
\begin{array}{l}
\bar{\mathbf{R}}_{1}[i] = \lambda\bar{\mathbf{R}}_{1}[i-1] + b[i-1]\mathbf{r}[i]\mathbf{r}^{H}[i]b^{\ast}[i-1]\\
\bar{\mathbf{R}}_{2}[i] = \lambda\bar{\mathbf{R}}_{2}[i-1] + b[i-2]\mathbf{r}[i]\mathbf{r}^{H}[i]b^{\ast}[i-2]\\
\bar{\mathbf{R}}_{3}[i] = \lambda\bar{\mathbf{R}}_{3}[i-1] + b[i-2]\mathbf{r}[i-1]\mathbf{r}^{H}[i-1]b^{\ast}[i-2]\\
\end{array}
\label{eq:Bidirect_LS_Autocorrelation}
\end{equation}
and the component cross-correlation vectors:
\begin{equation}
\begin{array}{l}
\bar{\mathbf{t}}_{1}[i] = \lambda\bar{\mathbf{t}}_{3}[i-1] + b[i-1]\mathbf{r}[i]\mathbf{r}^{H}[i-1]\mathbf{w}[i-1]b^{\ast}[i]\\
\bar{\mathbf{t}}_{2}[i] = \lambda\bar{\mathbf{t}}_{2}[i-1] + b[i-2]\mathbf{r}[i]\mathbf{r}^{H}[i-2]\mathbf{w}[i-1]b^{\ast}[i]\\
\bar{\mathbf{t}}_{3}[i] = \lambda\bar{\mathbf{t}}_{3}[i-1] + b[i-2]\mathbf{r}[i-1]\mathbf{r}^{H}[i-2]\mathbf{w}[i-1]b^{\ast}[i-1]\\
\end{array}.
\label{eq:Bidirect_LS_Crosscorrelation}
\end{equation}
The overall correlation structure is then formed from the summation
of the preceding expressions, yielding
\begin{equation}
\bar{\mathbf{R}}[i] = \bar{\mathbf{R}}_{1}[i] +
\bar{\mathbf{R}}_{2}[i] + \bar{\mathbf{R}}_{3}[i]
\end{equation}
and
\begin{equation}
\bar{\mathbf{t}}[i] = \bar{\mathbf{t}}_{1}[i] +
\bar{\mathbf{t}}_{2}[i] + \bar{\mathbf{t}}_{3}[i].
\end{equation}
where
\begin{equation}
\mathbf{w}[i] = \bar{\mathbf{R}}^{-1}[i]\bar{\mathbf{t}}[i].
\label{eq:Bidirect_LS}
\end{equation}
Similarly to the NLMS algorithm, performance improvements can be
expected if the variable switching and mixing factors,
(\ref{eq:swithcing_factors}) and (\ref{eq:mixing_factors}), are
incorporated into the correlation expressions. The resulting
expressions are
\begin{equation}
\mathbf{R}[i] = \rho_{1}[i]\mathbf{R}_{1}[i] + \rho_{2}[i]\mathbf{R}_{2}[i] + \rho_{3}[i]\mathbf{R}_{3}[i]
\label{eq:mixing_ls_auto}
\end{equation}
\begin{equation}
\mathbf{t}[i] = \rho_{1}[i]\mathbf{t}_{1}[i] + \rho_{2}[i]\mathbf{t}_{2}[i] + \rho_{3}[i]\mathbf{t}_{3}[i]
\label{eq:mixing_ls_cross}
\end{equation}
Introducing the above expression into the RLS framework would lead
to a low-complexity algorithm with improved convergence and
robustness compared to the NLMS of Section \ref{5_4_1_Stochastic
Gradient}. This requires the integration of
(\ref{eq:Bidirect_LS_Autocorrelation}) with the matrix inversion
lemma \cite{adapt_filt_Haykin, {adapt_filt_Diniz}}. However, the
derivation requires an expression with a rank-$1$ update of the form
\begin{equation}
\mathbf{R}[i] = \mathbf{R}[i-1] + \lambda \mathbf{r}[i]\mathbf{r}^{H}[i]
\end{equation}
for the autocorrelation matrix; a form which
(\ref{eq:Bidirect_LS_Autocorrelation}) is unable to fit into without
assumptions that cause a significant performance degradation. Consequently, an
alternative low-complexity algorithm to implement the LS solution given by
(\ref{eq:Bidirect_LS_Autocorrelation}) - (\ref{eq:Bidirect_LS}) is required.

\subsection{Conjugate Gradient Algorithm}
\label{5_4_3_Conjugate Gradient}

Due to the particular form of the bidirectional LS formulation and
the conventional RLS recursion, an alternative low-complexity method
is now derived. The CG technique has been chosen to avoid matrix
inversions and due to its excellent convergence properties
\cite{matrix_analysis_meyer,linear_non_linear_prog_luenburger,beamtrack}.
We begin the derivation of the proposed CG type algorithm with the
autocorrelation (\ref{eq:Bidirect_LS_Autocorrelation}) and
cross-correlation (\ref{eq:Bidirect_LS_Crosscorrelation}) structures
of subsection \ref{5_4_2_Least Squares}. Inserting them into the
standard CG quadratic form yields
\begin{equation}
J(\mathbf{w}) =\mathbf{w}^{H}[i]\mathbf{R}[i]\mathbf{w}[i]
-\mathbf{t}^{H}[i]\mathbf{w}[i]. \label{eq:cg_quadratic_bidirect}
\end{equation}
From \cite{linear_non_linear_prog_luenburger}, the unique minimiser of (\ref{eq:cg_quadratic_bidirect}) is also the minimiser of
\begin{equation}
\mathbf{R}[i]\mathbf{w}[i]=\mathbf{t}[i].
\end{equation}
This shows the suitability of the CG algorithm to the bidirectional problem. At
each time instant a number of iterations of the following method are required
to reach an accurate solution, where the iterations are indexed with the
variable $j$. Other single iteration CG methods are available but these depend
upon degeneracy - a term that describes the situation where the successive CG
vectors are not orthogonal \cite{cg_analysis_chang}. Consequently, we employ
the conventional method to ensure satisfactory convergence. At the
$i^{\mathrm{th}}$ time instant the gradient and direction vectors are
initialized as
\begin{equation}
\mathbf{g}_{0}[i]=\nabla_{\mathbf{w}[i]}C_{\rm
LS}(\mathbf{w}[i])=\mathbf{R}[i]\mathbf{w}_{0}[i]-\mathbf{t}[i]
\end{equation}
and
\begin{equation}
\mathbf{d}_{0}[i] = -\mathbf{g}_{0}[i],
\end{equation}
respectively, where the gradient expression is equivalent to those
used in the derivation of the previous algorithms. The vectors
$\mathbf{d}_{j}[i]$ and $\mathbf{d}_{j+1}[i]$ are $\mathbf{R}[i]$
orthogonal with respect to $\mathbf{R}[i]$ such that
$\mathbf{d}_{j}[i]\mathbf{R}[i]\mathbf{d}_{l}[i]=0$ for $j\neq l$.
At each iteration, the receive filter is updated as
\begin{equation}
\mathbf{w}_{j+1}[i] = \mathbf{w}_{j}[i] + \alpha_{j}[i]\mathbf{d}_{j}[i]
\label{eq:CG_filter_update}
\end{equation}
where $\alpha_{j}[i]$ is the {  minimizer} of $J(\mathbf{w}_{j+1}[i])$ such
that
\begin{equation}
\alpha_{j} = \frac{-\mathbf{d}_{j}^{H}\mathbf{g}_{j}[i]}{\mathbf{d}_{j}^{H}[i]\mathbf{R}[i]\mathbf{d}_{j}[i]}.
\end{equation}
The gradient vector is then updated according to
\begin{equation}
\mathbf{g}_{j+1}[i]=\mathbf{R}[i]\mathbf{w}_{j}[i]-\mathbf{t}[i]
\end{equation}
and a new conjugate gradient direction vector is obtained as given by
\begin{equation}
\mathbf{d}_{j+1}[i] = -\mathbf{g}_{j+1}[i] + \beta_{j}[i]\mathbf{d}_{j}[i]
\end{equation}
where
\begin{equation}
\beta_{j}[i] = \frac{\mathbf{g}_{j+1}^{H}[i]\mathbf{R}[i]\mathbf{d}_{j}[i]}{\mathbf{d}_{j}^{H}[i]\mathbf{R}[i]\mathbf{d}_{j}[i]}
\label{eq:CG_beta_update}
\end{equation}
ensures the $\mathbf{R}[i]$ orthogonality between
$\mathbf{d}_{j}[i]$ and $\mathbf{d}_{l}[i]$ where $j\neq l$. The
iterations (\ref{eq:CG_filter_update}) - (\ref{eq:CG_beta_update})
are then repeated until $j = j_{max}$.

The variable switching and mixing factors can be incorporated into
the algorithm to improve performance. This is achieved by operating
the CG algorithm over the modified correlation structures given by
(\ref{eq:mixing_ls_auto}) and (\ref{eq:mixing_ls_cross}).

\section{Analysis}
\label{5_5_Analysis}

{In this section, we analyze the proposed bidirectional algorithms to gain
insight of the expected performance but also to obtain further knowledge into
the operation of the proposed and existing algorithms. The unconventional form
of the proposed cost functions precludes the application of standard MSE
analysis. Consequently, we concentrate on the signal-to-interference-plus-noise
ratio (SINR) of the proposed algorithms in order to analyze their interference
suppression and tracking performance. We firstly study the NLMS algorithm and
the features of its weight error correlation matrix in order to arrive at an
analytical SINR expression. Following this, we explore the analogy between the
form of the bidirectional expression and convex combinations of adaptive
receive filters \cite{Bershad_transient_affine_combintation_2_lms,
MSE_performance_convex_comb_Garcia}.}

\subsection{SINR Analysis}
\label{5_5_1_SINR_Analysis}

{  Let us define the instantaneous SINR expression given by
\begin{equation}
\mathrm{SINR}_{\rm inst} \triangleq
\frac{\mathbf{w}^{H}[i]\mathbf{R}_{\mathrm{S}}
\mathbf{w}[i]}{\mathbf{w}^{H}[i]\mathbf{R}_{\mathrm{I}}\mathbf{w}[i]},
\label{sinr_inst}
\end{equation}}
where $\mathbf{R}_{S}$ and $\mathbf{R}_{I}$ are the signal and
interference and noise correlation matrices, into a form amenable to
analysis.

{  The receive filter error weight vector is described by
\begin{equation}
{\boldsymbol \varepsilon}[i] = \mathbf{w}[i]-\mathbf{w}_{o}[i],
\end{equation}
where $\mathbf{w}_{o}$ is the instantaneous standard MMSE receiver.}

{ Let us now describe the expression of the numerator of (\ref{sinr_inst}) with
the desired signal component:
\begin{equation}
\mathrm{S}_{\rm c} =
\boldsymbol{\varepsilon}^{H}[i]\mathbf{R}_{\mathrm{S}}\boldsymbol{\varepsilon}[i]
+ \boldsymbol{\varepsilon}^{H}[i]\mathbf{R}_{\mathrm{S}}\mathbf{w}_{o}[i]+
\overbrace{\mathbf{w}_{o}^{H}[i]\mathbf{R}_{\mathrm{S}}\mathbf{w}_{o}[i]}^{P_{\mathrm{S,opt}}[i]}+
\mathbf{w}_{o}^{H}[i]\mathbf{R}_{\mathrm{S}}\boldsymbol{\varepsilon}[i],
\label{sig_c}
\end{equation}
and of the interference plus noise component described by
\begin{equation}
\mathrm{S}_{i+n}= \boldsymbol{\varepsilon}^{H}[i]\mathbf{R}_{\mathrm{I}}
\boldsymbol{\varepsilon}[i]+ \boldsymbol{\varepsilon}^{H}[i]
\mathbf{R}_{\mathrm{I}}\mathbf{w}_{o}[i] + \underbrace{
\mathbf{w}_{o}^{H}[i]\mathbf{R}_{\mathrm{I}}
\mathbf{w}_{o}[i]}_{P_{\mathrm{I,opt}}[i]}+ {\mathbf w}_{o}^{H}[i] {\mathbf
R}_{\mathrm{I}}{\boldsymbol \varepsilon}[i]. \label{sig_in}
\end{equation}
} {  Taking the expectation and the trace of $\mathrm{S}_{c}$ and
$\mathrm{S}_{i+n}$, defining $\mathbf{K}[i] =
E[\boldsymbol{\varepsilon}[i]\boldsymbol{\varepsilon}^{H}[i]]$ and
$\mathbf{G}[i]=E[\mathbf{w}_{o}[i]\boldsymbol{\varepsilon}^{H}[i]]$, we can
define the following SINR expression:
\begin{equation}
\begin{split}
\mathrm{SINR}& \triangleq \frac{{\rm tr}\{E[S_c]\}}{{\rm tr}\{E[S_{i+n}]\}}\\ &
=\frac{{\rm tr}[\mathbf{K}[i]\mathbf{R}_{\mathrm{S}}+
\mathbf{G}[i]\mathbf{R}_{\mathrm{S}}+ P_{\mathrm{S,opt}}[i]+
\mathbf{G}^{H}[i]\mathbf{R}_{\mathrm{S}}]} {{\rm
tr}[\mathbf{K}[i]\mathbf{R}_{\mathrm{I}}+ \mathbf{G}[i]\mathbf{R}_{\mathrm{I}}+
P_{\mathrm{I,opt}}[i]+ \mathbf{G}^{H}[i]\mathbf{R}_{\mathrm{I}}]}.
\label{eq:SINR_Analysis_2}
\end{split}
\end{equation}
From (\ref{eq:SINR_Analysis_2}) it is clear that we need to pursue expressions
for $\mathbf{K}[i]$ and $\mathbf{G}[i]$ in order to reach an analytical
interpretation of the bidirectional NLMS scheme.}

Substituting the filter error weight vector into the filter update
expression of (\ref{eq:NLMS_Update}) yields a recursive expression
for the receive filter error weight vector described by
\begin{equation}
\begin{array}{ll}
\boldsymbol{\varepsilon}[i]&=\boldsymbol{\varepsilon}[i-1]\\ &
+\left[\mathbf{I}+\mu\mathbf{r}[i]b[i-1]\mathbf{r}^{H}[i-1]b^{\ast}[i]
-\mu\mathbf{r}[i]b[i-1]\mathbf{r}^{H}[n]b^{\ast}[i-1]\right.\\
&+\mu\mathbf{r}[i]b[i-2]\mathbf{r}^{H}[i-2]b^{\ast}[i]-
\mu\mathbf{r}[i]b[i-2]\mathbf{r}^{H}[n]b^{\ast}[i-2]\\
&\left.+\mu\mathbf{r}[i-1]b[i-2]\mathbf{r}^{H}[i-2]b^{\ast}[i-1]
-\mu\mathbf{r}[i-1]b[i-2]\mathbf{r}^{H}[i-1]b^{\ast}[i-2]\right]\boldsymbol{\varepsilon}[i-1]\\
&+\mu\mathbf{r}[i]b[i-1]e^{\ast}_{o,1}[i]\\
&+\mu\mathbf{r}[i]b[i-2]e^{\ast}_{o,2}[i]\\
&+\mu\mathbf{r}[i-1]b[i-2]e^{\ast}_{o,3}[i]\\
\end{array},
\label{eq:bidirectional_stochastic_diff}
\end{equation}
where the terms $e_{o,1-3}$ denote the error terms of
(\ref{eq:NMLS_Gradient_Upadted_w}) when the optimum filter $\mathbf{w}_{o}$ is
used. {  Utilizing the direct averaging approach developed by Kushner
\cite{kushner_weak_convergence}, often invoked in this type of stochastic
analysis, the solution to the stochastic difference equation of
(\ref{eq:bidirectional_stochastic_diff}) can be approximated by the solution to
a second equation \cite{adapt_filt_Haykin, sayed_adaptive_filters}, such that}
\begin{equation}
\begin{array}{l}
E\left[\mathbf{I} + \mu\mathbf{r}[i]b[i-1]\mathbf{r}^{H}[i-1]b^{\ast}[i]
-\mu\mathbf{r}[i]b[i-1]\mathbf{r}^{H}[n]b^{\ast}[i-1]\right.\\
+\mu\mathbf{r}[i]b[i-2]\mathbf{r}^{H}[i-2]b^{\ast}[i]
-\mu\mathbf{r}[i]b[i-2]\mathbf{r}^{H}[n]b^{\ast}[i-2]\\ \left.
+\mu\mathbf{r}[i-1]b[i-2]\mathbf{r}^{H}[i-2]b^{\ast}[i-1]
-\mu\mathbf{r}[i-1]b[i-2]\mathbf{r}^{H}[i-1]b^{\ast}[i-2]\right]\\
=\\
\mathbf{I}+\mu\mathbf{F}_{1}-\mu\mathbf{R}_{1}+\mu\mathbf{F}_{2}-\mu\mathbf{R}_{2} +\mu\mathbf{F}_{3}-\mu\mathbf{R}_{3}
\end{array},
\label{eq:direct_ave_nlms_analysis}
\end{equation}
where $\mathbf{F}$ and $\mathbf{R}$ are correlations matrices.
Specifically, $\mathbf{R}_{1-3}$ are autocorrelation matrices given
by
\begin{equation}
\begin{array}{l}
\mathbf{R}_{1} = E\left[\mu\mathbf{r}[i]b^{\ast}[i-1]\mathbf{r}^{H}[i]b^{\ast}[i-1]\right]\\
\mathbf{R}_{2} = E\left[\mu\mathbf{r}[i]b^{\ast}[i-2]\mathbf{r}^{H}[i]b^{\ast}[i-2]\right]\\
\mathbf{R}_{3} = E\left[\mu\mathbf{r}[i-1]b^{\ast}[i-2]\mathbf{r}^{H}[i-1]b^{\ast}[i-1]\right]\\
\end{array}
\end{equation}
and $\mathbf{F}_{1-3}$ cross-time instant correlation matrices,
given by
\begin{equation}
\begin{array}{l}
\mathbf{F}_{1} = E\left[\mu\mathbf{r}[i]b^{\ast}[i-1]\mathbf{r}^{H}[i-1]b^{\ast}[i]\right]\\
\mathbf{F}_{2} = E\left[\mu\mathbf{r}[i]b^{\ast}[i-2]\mathbf{r}^{H}[i-2]b^{\ast}[i]\right]\\
\mathbf{F}_{3} = E\left[\mu\mathbf{r}[i-1]b^{\ast}[i-2]\mathbf{r}^{H}[i-2]b^{\ast}[i-1]\right].\\
\end{array}
\end{equation}
Using (\ref{eq:direct_ave_nlms_analysis}) and the independence
assumption of
$E\left[e_{o,1-3}[i]\boldsymbol{\varepsilon}[i]\right]=0$,
$E\left[\mathbf{r}^{H}[i]\mathbf{r}[i-1]\right]=0$ and
$E\left[b_{k}[i]b_{k}[i-1]\right]=0$, we arrive at the expression
for $\mathbf{K}[i]$
\begin{equation}
\begin{array}{ll}
\mathbf{K}[i]=&\left[\mathbf{I}+\mu\mathbf{F}_{1}-\mu\mathbf{R}_{1}
+\mu\mathbf{F}_{2}-\mu\mathbf{R}_{2} +\mu\mathbf{F}_{3} -\mu\mathbf{R}_{3}
\right]  \mathbf{K}[i-1] \left[\mathbf{I}+\mu\mathbf{F}_{1}-\mu\mathbf{R}_{1}
+\mu\mathbf{F}_{2}-\mu\mathbf{R}_{2} +\mu\mathbf{F}_{3}
-\mu\mathbf{R}_{3} \right] \\
&+ \mu^{2}\mathbf{R}_{1}J_{min,1}[i] \\
&+ \mu^{2}\mathbf{R}_{2}J_{min,2}[i] \\
&+ \mu^{2}\mathbf{R}_{1}J_{min,3}[i] \\
\end{array}
\end{equation}
where $J_{{\rm min},j}[i] = |e_{o,j}|^{2}$. Following a similar method, an
expression for $\mathbf{G}[i]$ can also be reached
\begin{equation}
\mathbf{G}[i] = \mathbf{G}[i-1]\left[\mu\mathbf{F}_{1} -
\mu\mathbf{R}_{1} + \mu\mathbf{F}_{2} - \mu\mathbf{R}_{2} +
\mu\mathbf{F}_{3} - \mu\mathbf{R}_{3}\right].
\end{equation}
At this point we study the derived expression to gain an insight
into the operation of the bidirectional algorithm and the origins of
its advantages over the conventional differential scheme. Equivalent
expressions for the conventional stochastic gradient scheme are
given by
\begin{equation}
\begin{array}{ll}
\mathbf{K}[i]=&\left[\mathbf{I}+\mu\mathbf{F}_{1}-\mu\mathbf{R}_{1} \right]\mathbf{K}[i-1]
\left[\mathbf{I}+\mu\mathbf{F}_{1}-\mu\mathbf{R}_{1}\right] \\
&+ \mu^{2}\mathbf{R}_{1}J_{min,1}[i] \\
\mathbf{G}[i] = &\mathbf{G}[i-1]\left[\mu\mathbf{F}_{1} - \mu\mathbf{R}_{1}\right].
\end{array}
\end{equation}
The bidirectional scheme has a number of additional correlation
terms compared to the conventional scheme. Evaluating the cross-time
instant matrices yields
\begin{equation}
\begin{array}{l}
\mathbf{F}_{1} = |a_{1}|^{2}\mathbf{c}_{1}\mathbf{c}_{1}^{H}\underbrace{E\left[h[i]h^{\ast}[i-1]\right]}_{f_{1}[i]}\\
\mathbf{F}_{2} = |a_{1}|^{2}\mathbf{c}_{1}\mathbf{c}_{1}^{H}\underbrace{E\left[h[i]h^{\ast}[i-2]\right]}_{f_{2}[i]}\\
\mathbf{F}_{3} =  |a_{1}|^{2}\mathbf{c}_{1}\mathbf{c}_{1}^{H}\underbrace{E\left[h[i-1]h^{\ast}[i-2]\right]}_{f_{3}[i]}\\
\end{array}.
\label{eq:F_expressions}
\end{equation}
From the expression above it is clear that the underlying factor
that governs the SINR performance of the algorithms is the
correlation between the considered time instants, $f_{1-3}$,
data-ruse and the use of $f_{2}$. Accordingly, it is the additional
correlation factors that the proposed bidirectional algorithms
possess that enhances its performance compared to the conventional
scheme, confirming the initial motivation behind the proposition of
the bidirectional approach. Lastly, the $f_{1-3}$ expressions of
(\ref{eq:F_expressions}) are the factors that influence the optimum
number of considered time instants.

\subsection{Combinations of Adaptive Receive Filters}

To further our understanding of the bidirectional algorithms we
follow a heuristic and complementary approach that leads to an
analogy with a combination of adaptive filters
\cite{Bershad_transient_affine_combintation_2_lms}. The
bidirectional LS solution given by (\ref{eq:Bidirect_LS}) is made up
of $6$ constituent correlation structures that result in a filter
output of
\begin{equation}
y[i] = \left[\left(\rho_{1}\mathbf{R}_{1}[i] +
\rho_{3}\mathbf{R}_{2}[i] +
\rho_{3}\mathbf{R}_{3}[i]\right)^{-1}\left(\rho_{1}\mathbf{t}_{1}[i]+
\rho_{2}\mathbf{t}_{2}[i] +
\rho_{3}\mathbf{t}_{3}[i]\right)\right]^{H} \mathbf{r}[i].
\end{equation}
Decomposing the expression above leads us to an expression where the
signal $y[i]$ is formed from the output of 3 individual adaptive
receive filters
\begin{equation}
\begin{array}{ll}
y[i] &=\left[\left(\mathbf{R}_{1}[i]+\frac{\rho_{2}}{\rho_{1}}\mathbf{R}_{2}[i]+\frac{\rho_{3}}{\rho_{1}}\mathbf{R}_{3}[i]\right)^{-1}\mathbf{t}_{1}[i]\right]^{H}\mathbf{r}[i]\\
&\\
&+\left[\left(\mathbf{R}_{1}[i]+\frac{\rho_{1}}{\rho_{2}}\mathbf{R}_{2}[i]+\frac{\rho_{3}}{\rho_{2}}\mathbf{R}_{3}[i]\right)^{-1}\mathbf{t}_{2}[i]\right]^{H}\mathbf{r}[i]\\
&\\
&+\left[\left(\mathbf{R}_{1}[i]+\frac{\rho_{1}}{\rho_{3}}\mathbf{R}_{2}[i]+\frac{\rho_{2}}{\rho_{3}}\mathbf{R}_{3}[i]\right)^{-1}\mathbf{t}_{3}[i]\right]^{H}\mathbf{r}[i]\\
\end{array}.
\end{equation}
{  This is equivalent to a convex combination of adaptive receive filters with
varying $\lambda$ \cite{Bershad_transient_affine_combintation_2_lms,
MSE_performance_convex_comb_Garcia}, where each of the 3 filters focuses on the
correlation between the 2 of the 3 considered time instants. However, the
presence of the autocorrelation matrices in the inverses of the expression also
indicates that the remaining time instants also influence the structure of each
filter. Although the mixing factors are not separable we can interpret them as
a form of weighting that is present in conventional combinations of adaptive
filters. This explains in part the additional control and performance they
provide.}

\section{Simulations}
\label{5_6_Simulations}

In this section, the proposed bidirectional adaptive algorithms are applied to
conventional multiuser and cooperative DS-CDMA systems using the signal models
described in Section \ref{models}. The application of the proposed algorithms
to multiple-antenna and multi-carrier systems is straightforward and requires a
change in the signal models. The individual Rayleigh fading channel
coefficients, $h[i]$, are generated using Clarke's model
\cite{W_Jakes_microwave_comms} where 20 scatterers are assumed. In all
simulations the number of packets is denoted by $N_{p}$ and the fading rate is
given by the dimensionless normalized fading parameter, $T_{s}f_{d}$, where
$T_{s}$ is the symbol period and $f_{d}$ is the Doppler frequency shift. The
convergence parameters of the algorithms have been optimized resulting in
step-sizes forgetting factors of 0.1 and 0.99, respectively,
$\lambda_{\mathrm{e}}=0.95$, $\lambda_{M}=0.99$ and the number of CG
iterations, $j_{\mathrm{max}}=5$.

As detailed in Section \ref{5_5_Analysis}, the proposed algorithms do not
minimize the same MSE as a conventional MMSE receiver; therefore, the MSE is
not an adequate performance metric. As a result, BER and SINR based metrics are
chosen for the purpose of comparison between existing algorithms and the
optimum MMSE solution. Due to the rapidly fading channel, the instantaneous
SNR, $\mathrm{SNR}_{\mathrm{i}}$, is highly variable and so the SINR alone is
also not a satisfactory metric. To overcome this it is normalized by the
instantaneous SNR to give $\frac{SINR}{SNR_{\mathrm{i}}}$. This value is
negative in all simulations and directly reflects the MUI interference
suppression and tracking capabilities of the proposed algorithms
\cite{CDMA_Serverly_Time_Varying_Channel_Madhow, Diff_MMSE_Madhow}.

\subsection{Conventional DS-CDMA}
\label{5_6_1_Conventional_DS-CDMA}

Here we apply the adaptive algorithms of Section \ref{5_4_Adaptive Algorithms}
to interference suppression in the uplink of a multiuser DS-CDMA system
described in Section \ref{models}. Each simulation is averaged over $N_{p}$
packets and detailed parameters are specified in each plot.

\subsubsection{Analytical Results}

\begin{figure}[!h]
\centering
\includegraphics[width=0.95\columnwidth]{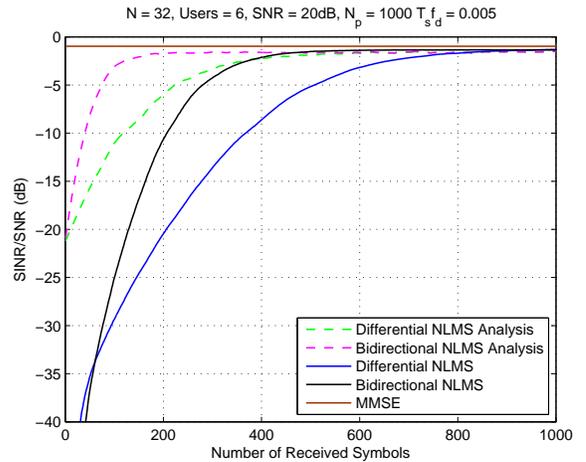}
\vspace{-0.5em} \caption{SINR performance comparison of simulated
and analytical proposed NLMS algorithms over a single path channel.}
\label{fig:SINRvsSym_Known_Correlation}
\end{figure}

We first assess the analytical expressions derived in Section
\ref{5_5_1_SINR_Analysis} and their agreement with simulated
results. Central to the performance of the differential and
bidirectional schemes are the correlation factors $f_{1-3}$ and the
related assumption of $h_{1}[i] \approx h_{1}[i-1]$. Examining the
effect of the fading rate on the value of $f_{1-3}$ shows that
$f_{1}\approx f_{2} \approx f_{3}$ at fading rates of up to
$T_{s}f_{d}=0.01$. Consequently, after a large number of received
symbols with high total receive power
\begin{equation}
3\left[\mathbf{I}+\mu\mathbf{F}_{1} -
\mu\mathbf{R}_{1}\right]\approx\left[\mathbf{I}+\mu\mathbf{F}_{1} -
\mu\mathbf{R}_{1} + \mu\mathbf{F}_{2} - \mu\mathbf{R}_{2} +
\mu\mathbf{F}_{3} - \mu\mathbf{R}_{3}\right],
\label{eq:diff_bi_equivalence}
\end{equation}
due to the decreasing significance of the identity  matrix.  This
indicates that the expected value of the SINR, of the bidirectional
scheme, once $f_{1}\approx f_{2} \approx f_{3}$ have stabilised,
should be similar to the differential scheme. A second implication
is that the bidirectional scheme should converge towards the MMSE
level due to the equivalence between the differential scheme and the
MMSE solution \cite{Diff_MMSE_Madhow}. Fig.
\ref{fig:SINRvsSym_Known_Correlation} illustrates the analytical
performance using the expressions given in Section
\ref{5_5_1_SINR_Analysis}.

The correlation matrices are calculated via ensemble averages prior to the
start of the algorithm and $\mathbf{G}[0]=\mathbf{K}[0]=\mathbf{I}$.  In Fig.
\ref{fig:SINRvsSym_Known_Correlation} one can see the convergence of the
simulated schemes to the analytical and MMSE plots, validating the presented
analysis. Due to the highly dynamic nature of the channel, using the expected
values of the correlation matrix alone cannot capture the true transient
performance of the algorithms. However, the convergence period of the
analytical plots within the first 200 iterations can be considered to be within
the coherence time and therefore give an indication of the transient
performance relative to other analytical plots. Using this justification and
the aforementioned analysis, advantages should be present in the transient
phase due to the additional correlation information supplied by
$\mathbf{F}_{2}$ and $\mathbf{F}_{3}$. This conclusion is supported by Fig.
\ref{fig:SINRvsSym_Known_Correlation} and the similar forms of the analytical
and simulated schemes relative to each other and their subsequent convergence.

\subsubsection{SINR Performance}

\begin{figure}[!h]
\centering
\includegraphics[width=0.95\columnwidth]{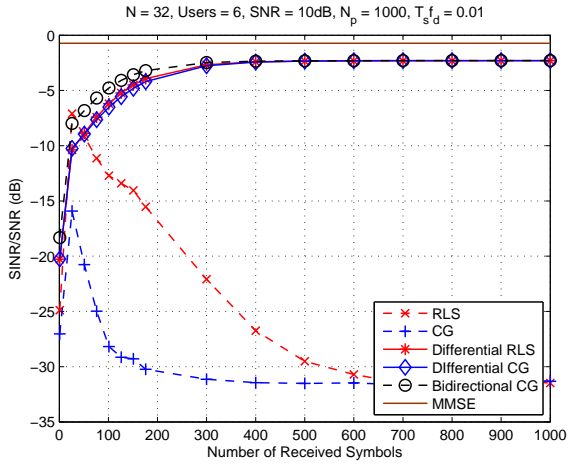}
\vspace{-0.5em} \caption{SINR performance comparison of proposed CG
algorithms over a single path channel where all schemes have been
trained with 150 symbols and then switched to decision directed
mode.} \label{fig:SINRvsSym_CG}
\end{figure}

\begin{figure}[!h]
\centering
\includegraphics[width=0.95\columnwidth]{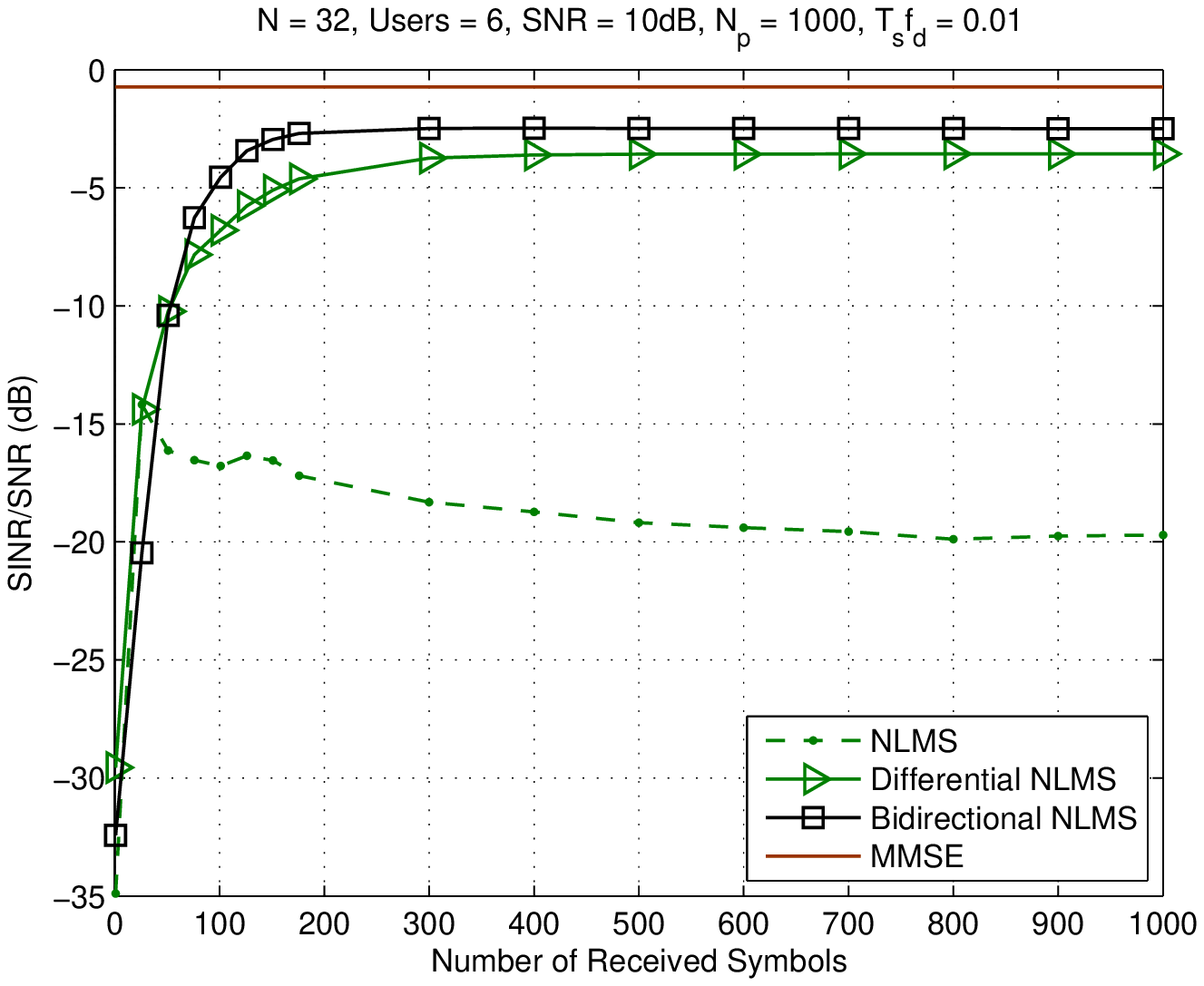}
\vspace{-0.5em} \caption{SINR performance comparison of proposed
NLMS algorithms over a single path channel where all schemes have
been trained with 150 symbols and then switched to decision directed
mode.} \label{fig:SINRvsSym_NLMS}
\end{figure}

The SINR/SNR performance of the proposed algorithms is given by
Figs. \ref{fig:SINRvsSym_CG} and \ref{fig:SINRvsSym_NLMS}. The
performance of the CG implementation of the differential algorithm
is marginally below that of the RLS during convergence but the
bidirectional scheme provides noticeable improvements. The
differential and bidirectional algorithms converge close to the MMSE
optimum as expected from the previous analysis. The bidirectional
NLMS algorithm provides more significant improvements over the
differential scheme, both in the final stages of convergence and
steady-state. These differences can be accounted for by the reduced
receive signal power; the matrices equivalent to $\mathbf{F}2$ and
$\mathbf{F}3$ improving the consistency of the steady-state
performance by reducing the impact of weakly-correlated samples; and
the NLMS's suitability to data reuse as in the affine projection
(AP) algorithm. As expected, the conventional adaptive schemes are
unable to converge or track the solution due to the more demanding
task of tracking both the fading coefficients and suppressing MUI.

\begin{figure}[!h]
\centering
\includegraphics[width=0.95\columnwidth]{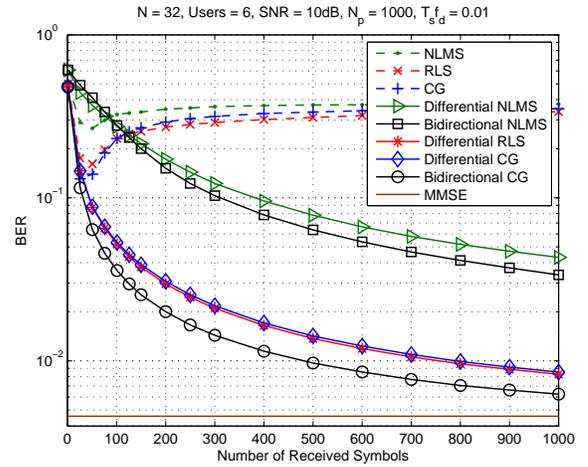}
\vspace{-0.5em} \caption{BER performance comparison of proposed
schemes during training over a single path channel.}
\label{fig:BERvsSym}
\end{figure}

The BER performance of the differential and bidirectional schemes is
illustrated in Fig. \ref{fig:BERvsSym}, where the system parameters are equal
to those of Figs. \ref{fig:SINRvsSym_CG} and \ref{fig:SINRvsSym_NLMS}. The RLS
and CG algorithms converge to near the MMSE level with the bidirectional scheme
providing a performance improvement over other considered algorithms. The NLMS
schemes exhibit slower BER convergence compared to their SINR performance but
reach a level where decision directed operation can take place in a severely
fading channel. Due to the superior performance of the CG and RLS based
algorithms we predominantly focus on their performance for the remainder of
this section.

\begin{figure}[!h]
\centering
\includegraphics[width=0.95\columnwidth]{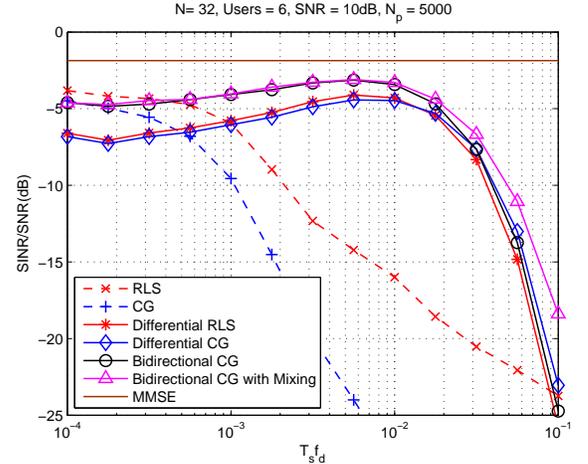}
\vspace{-0.5em} \caption{SINR performance versus fading rate of the
proposed CG schemes over a single path channel after 200 training
symbols.} \label{fig:SINRvsFading_CG}
\end{figure}

Fig. \ref{fig:SINRvsFading_CG} illustrates the performance of the
proposed CG and RLS algorithms as the fading rate is increased. The
conventional schemes with RLS and CG algorithms are unable to cope
with fading rates in excess of $T_{s}f_{d}=0.005$  and begin to
diverge at the completion of the training sequence. The proposed
bidirectional scheme outperforms the differential schemes but the
performance begins to decline once fading rates above
$f_{d}T_{s}=0.01$ are reached. Once again, the increase in
performance of the bidirectional scheme can be accounted for by the
increased correlation information supplied by the matrices
$\mathbf{F}_2$ and $\mathbf{F}_{3}$ and data reuse. The introduction
of the mixing factors into the bidirectional algorithm improves
performance further, especially at higher fading rates. A first
reason for this is the improvement in consistency as previously
mentioned. However, a second more significant reason can be
established by referring back to the observations on the correlation
factors $f_{1-3}$. Although fading rates of $0.01$ may be fast, the
assumption $h[i-2]\approx h[i-1]\approx h[i]$ is still valid.
Consequently, $f_{1}\approx f_{2} \approx f_{3}$ and equal weighting
is adequate. However, as the fading rate increases beyond
$T_{s}f_{d}=0.01$ this assumption breaks down and the correlation
information requires unequal weighting for optimum performance, a
task fulfilled by the adaptive mixing factor.

\begin{figure}[!h]
\centering
\includegraphics[width=0.95\columnwidth]{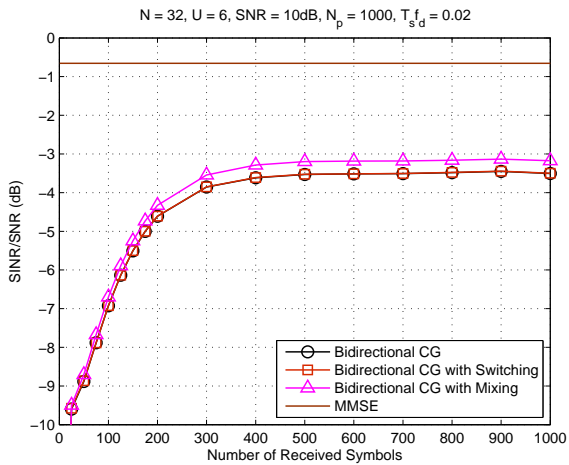}
\vspace{-0.5em} \caption{SINR performance over a single path channel
of the proposed schemes with switching and mixing factors.}
\label{fig:SINRvsSym_Switching_CG}
\end{figure}

A more detailed plot illustrating the performance advantages of the
CG switching and mixing parameters presented in is given by Fig.
\ref{fig:SINRvsSym_Switching_CG}. The switching approach provides
little improvement over the standard bidirectional scheme due to its
discrete and non-adaptive operation. As previously covered, a low
instantaneous value of $f_{1-3}$, as indicated by a large power
differential, does not indicate that all information gathered on
$f_{1-3}$ is redundant. The mixing parameter implementations address
this shortcoming by adaptively setting the parameters via the error
weight expression (\ref{eq:mixing_factors}) that accurately reflects
the averaged correlation factors. At a fading rate of
$T_{s}f_{d}=0.02$ the assumption of $f_{1}\approx f_{2}\approx
f_{3}$ begins to diminish in accuracy and therefore unequal
weighting is required for performance in excess of the standard
bidirectional scheme, as previously mentioned and shown in Fig.
\ref{fig:SINRvsSym_Switching_CG}.

%

\begin{figure}[!h]
\centering
\includegraphics[width=0.95\columnwidth]{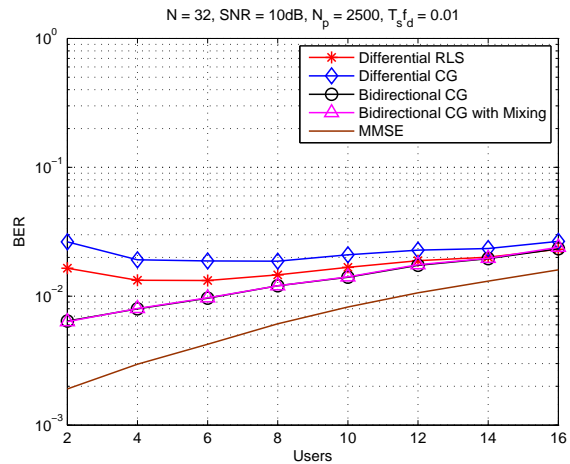}
\vspace{-0.5em}
\caption{BER performance against system loading
after 500 symbols of the proposed schemes over a single path
channel. Schemes are trained with 150 symbols and then switch to
decision directed operation.} \label{fig:BERvsUsers}
\end{figure}

The MUI suppression of the proposed and existing schemes is given by Fig.
\ref{fig:BERvsUsers}. The bidirectional scheme has significantly improved
multiuser performance compared to the differential algorithms at low system
loads but diminishes as the number of users increases. This behavior supports
the analytical conclusions of Section \ref{5_5_1_SINR_Analysis} by virtue of
the convergence of the differential and bidirectional schemes and the
increasing accuracy of (\ref{eq:diff_bi_equivalence}) as system loading, and
therefore received power, increases.

\subsection{Cooperative DS-CDMA}
\label{5_6_2_Cooperative_DS-CDMA}

To further demonstrate the performance of the proposed schemes in cooperative
relaying systems \cite{tds}, we apply them to an AF cooperative DS-CDMA system
detailed in Section \ref{models}.

\begin{figure}
\centering
 \includegraphics[width=0.95\columnwidth]{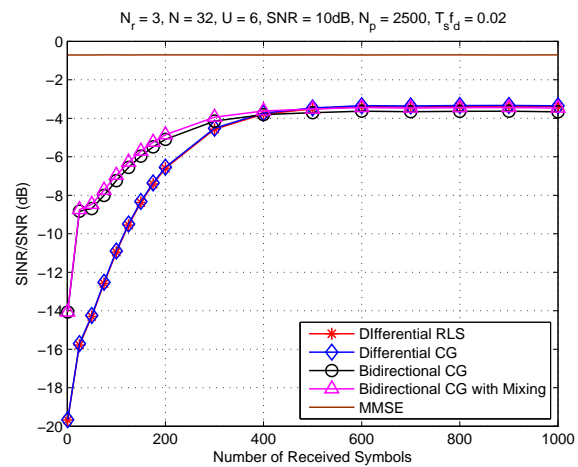}
 \vspace{-0.5em}
  \caption{SINR performance of the proposed CG schemes during training in a single path cooperative DS-CDMA system.}
  \label{fig:SINR_Cooperative}
\end{figure}

Fig. \ref{fig:SINR_Cooperative} shows that the bidirectional scheme
obtains performance benefits over the differential schemes during
convergence but, as expected, the performance gap closes as
steady-sate is reached. The inclusion of variable mixing parameters
improves performance but to a lesser extent than non-cooperative
networks due to the more challenging scenario of compounding highly
time-variant channels.

\begin{figure}
\centering
 \includegraphics[width=0.95\columnwidth]{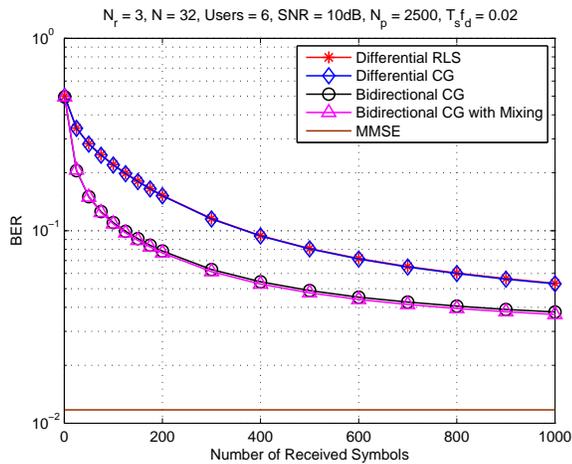}
 \vspace{-0.5em}
  \caption{BER performance of the proposed CG schemes in a single path cooperative DS-CDMA system.}
  \label{fig:BER_Cooperative}
\end{figure}

The improvement BER brought about by the bidirectional schemes is
evident from Figure \ref{fig:BER_Cooperative}. However, the more
challenging environment of a cooperative system with compounded
rapid fading has impacted on the BER performance of the schemes, as
evidenced by the increased performance gap between the proposed
schemes and MMSE reception.

\section{Conclusions}
\label{5_7_Conclusions}

In this paper, we have presented a bidirectional MMSE framework that
exploits the correlation characteristics of rapidly varying fading
channels to overcome the problems associated with conventional
adaptive interference suppression techniques in such channels. 
An analysis of the proposed schemes has been performed and the
reasons behind the performance improvements shown to be the
additional correlation information, data reuse and optimised
correlation factor weighting. The conditions under which the
differential and bidirectional schemes are equivalent have also been
established and the steady-state implications of this detailed.
Finally, the proposed algorithms have been assessed in standard and
cooperative multiuser DS-CDMA systems and shown to outperform both
differential and conventional schemes.

\bibliographystyle{IEEEtran}

\bibliography{bid_alg_rev}
\end{document}